\RequirePackage{colortbl} 
\documentclass[12pt]{article}
\usepackage{authblk}
\usepackage{url}
\usepackage[utf8]{inputenc}
\usepackage{ae,aecompl}
\usepackage{epsfig}
\usepackage{amsmath,amssymb}
\usepackage[T1]{fontenc} 
\usepackage{bm}
\usepackage{amsmath}
\usepackage{setspace}
\usepackage{algorithm}
\usepackage{algpseudocode}
\usepackage[dvipsnames]{xcolor}
\usepackage{mathtools}
\usepackage{geometry}
\def\diag{\mathop{\rm diag}\nolimits}

\def\qss{\mathbf{C}(:,1)}

\def\qsse{{\mathbf{\check{c}}}}
\def\Qsse{\mathbf{\check{C}}}

\def\nie{\check{n}_i}

\def\ne{\check{n}}
\def\Ne{\check{N}}

\def\Nee{\bar{N}}
\def\nee{\bar{n}}

\def\nc{\check{n}}

\def\trace{\mathop{\rm trace}\nolimits}

\def\Fi{\mathcal{F}_i}
\def\iFi{\mathcal{F}_i^{-1}}
\def\Fd{\mathcal{F}^{[d]}}
\def\iFd{\mathcal{F}^{[-d]}}
\newcommand{\vek}[1]{\mathchoice{\displaystyle\boldsymbol#1}
{\textstyle\boldsymbol#1}{\scriptstyle\boldsymbol#1}
{\scriptscriptstyle\boldsymbol#1}}

\def\sinc{\mbox{sinc}}

\newcommand{\refsec}[1]{Section~\ref{#1}}
\newcommand{\refsecp}[1]{(Section~\ref{#1})}

\newcommand{\BIGOP}[1]{\mathop{\mathchoice%
{\raise-0.22em\hbox{\huge $#1$}} {\raise-0.05em\hbox{\Large $#1$}}
{\hbox{\large $#1$}}{#1}}}

\newcommand{\BIGboxplus}{\mathop{\mathchoice%
{\raise-0.35em\hbox{\huge $\boxplus$}}%
{\raise-0.15em\hbox{\Large $\boxplus$}}{\hbox{\large
$\boxplus$}}{\boxplus}}}

\newcommand{\xib}{\bm{\xi}}
\newcommand{\lambdab}{\bm{\lambda}}

\title{Kriging in Tensor Train data format}

\author[1]{Sergey Dolgov}
\author[2]{Alexander Litvinenko}
\author[3]{Dishi Liu}

\affil[1]{
  University of Bath\\
Claverton Down, Bath, BA2 7AY, United Kingdom\\
  e-mail: s.dolgov@bath.ac.uk}
   
\affil[2]{RWTH Aachen \\
  Kackertstr. 9C, 52072, Aachen, Germany\\
  e-mail: litvinenko@uq.rwth-aachen.de}
\affil[3]{
  Institute of Scientific Computing,
Technische Universit\"at Braunschweig\\
  M\"uhlenpfordtstrasse 23,
D-38106 Braunschweig, Germany\\
  e-mail: d.liu@tu-bs.de }

\begin{document}

\maketitle

\begin{abstract}
Combination of low-tensor rank techniques and the Fast Fourier transform (FFT) based methods had turned out to be prominent in accelerating various statistical operations such as Kriging, computing conditional covariance, geostatistical optimal design, and others. However, the approximation of a full tensor by its low-rank format can be computationally formidable. In this work, we incorporate the robust  Tensor Train (TT) approximation of covariance matrices and the efficient TT-Cross algorithm into the FFT-based Kriging. It is shown that
here the computational complexity of Kriging is reduced to $\mathcal{O}(d r^3 n)$,
where $n$ is the mode size of the estimation grid, $d$ is the number of variables (the dimension),
and $r$ is the rank of the TT approximation of the covariance matrix. For many popular covariance functions
the TT rank $r$ remains stable for increasing $n$ and $d$. The advantages of this approach against those using plain
FFT are demonstrated in synthetic and real data examples.
\end{abstract}

\textbf{This work is dedicated to our wonderful colleague Prof. Hermann G.
Matthies on the occasion of his 68th birth anniversary.}

\tableofcontents
\section{Introduction\label{sec:introduction}}
Kriging is an interpolation method that makes estimates of unmeasured quantities based on (sparse) scattered measurements. It is widely applied in the estimation of some spatially distributed quantities such as daily moisture, rainfall intensities, temperatures, contaminant 
concentrations or hydraulic conductivities, etc. \cite{Matheron_1971,Journel_Huijbregts_1978_book}. Kriging is also used as a surrogate of some complex physical models for the purpose of efficient uncertainty quantification (UQ), in which it estimates the model response under some random perturbation of the parameters.  In the first case the  estimation grids are usually in two or three dimensions \cite{Wesson_Pegram_2004,finke2004mapping,haylock2008european} or four dimensions in a space-time Kriging
\cite{bogaert1996comparison,Kyriakidis_Journel_1999MG_Space_Time_Kriging,de2011toward}, while in the latter the dimension number could be much larger (equals to the number of uncertain parameters). 
When considering finely resolved estimation grids (which is often the case for UQ jobs), 
Kriging can easily exceed the computational capacity of modern computers. In this case estimation variance of Kriging or solving the related geostatistical optimal design problems incurs even higher computational costs  
\cite{Mueller_2007_Spatial_Data,Nowak_2010MG_Measures_Uncertainty,Spoeck_Pilz_2010SERRA_OD_convex}.
Kriging mainly involves three computational tasks. The first is solving a $N\times N$ system of equations to obtain the
Kriging weights, where $N$ is the number of measurements. Despite its $\mathcal{O}(N^3)$ complexity this task is better manageable since $N$ is usually much smaller than the number of estimates on a fine grid, $\Nee=\nee^d$, $d$ the dimensionality, especially when the measurement is expensive like for complex physical models. The second task is to compute the $\Nee$ Kriging estimates by multiplying the  weights vector to the $\Nee\times N$ cross-covariance
matrix between measurements and unknowns. The third task is to evaluate the $\bar  N$ estimation
variances as the diagonal of a $\bar  N\times \bar  N$ conditional covariance matrix. If we take the optimal design of sampling into account, there is an additional task 
to repeatedly evaluate the $\Nee \times \Nee$ conditional covariance matrix for the purpose of a high-dimensional non-linear optimization
\cite{Kollat_al_2008AWR_epsilon_hBOA_MOO,Shah_Reed_2011OR_MOO_comparison_knapsack,Reed_Minsker_Valocchi_2000WRR_Monitoring_Genetic_Algorithm}.

Remarkable progress had been made in speeding up Kriging computations by Fast Fourier transform (FFT) \cite{Fritz_Nowak_Neuweiler_2009_FFT_Kriging}. The low-rank tensor decomposition techniques brought a further possible reduction in the time cost, since $d$-dimensional FFT on a tensor in low-rank format can be made at the cost of a series of 1-dimensional FFT's, as exemplified in \cite{Vondrejc2019} by using canonical, Tucker and Tensor Train formats of tensors. The work in \cite{NowakLitv2013} brought a significant further reduction of computational cost for the second and third Kriging tasks as well as the task for the optimal design of sampling by applying a low-rank canonical tensor approximation to the vectors of interest.

In this paper, we enhance the methodology proposed in \cite{NowakLitv2013} by
 employing a more robust low-rank Tensor Train (TT) format instead of the canonical format. We apply the TT-cross algorithm for efficient approximation of tensors, which is a key improvement compared to the method introduced in \cite{NowakLitv2013} where the low-rank format of the covariance matrix was assumed to be given. We also consider a more broad Mat\'ern class of covariance functions.

The current work improves the applicability of the use of low-rank techniques in the FFT-based Kriging. We achieve a reduction of the computational complexity of Kriging to the level of $\mathcal{O}(dr^3  \bar n )$, where
$r$ is the considered TT rank of the approximation, and
$\bar n$ is the number of grid points in \emph{one} direction, such that $\bar  N=\bar n^d$ is the total number of estimated points.

We assume second-order stationarity for the covariance function and simple Kriging on 
a rectangular, equispaced grid parallel to the axes.

We also discuss possible extensions to non-rectangular domains and to general (scattered) measurement points. In such cases, the tensor ranks may significantly increase, up to the full rank.
For the cases when FFT technique is not applicable the authors of  \cite{Saibaba_Kitanidis_2012WRR_HMatrix_QLGA,LitvGenton2019,Litv17_HLIBPRO,Litv08_KLE} applied
the hierarchical matrix technique
($\mathcal{H}$-matrices).
A parallel implementation of Kriging was done in \cite{pesquer2011parallel}.

\subsection{State of the art for FFT-based Kriging}

Let us assume that the covariance function is second-order stationary and   is discretized on a tensor (regular and equispaced grid) mesh with $\bar  N=\bar  n^d$
points. Then the $\bar  N \times \bar  N$ auto-covariance
matrix of the unknowns has a symmetric (block-) Toeplitz structure
\refsecp{Sub:Toeplitz}, which can be extended to a (block-) circulant matrix by a periodic embedding in which the number of rows and columns is enlarged, for example,  from $\bar  N$ to $\check N=2\bar  N+1$ \cite{Pegram_2004,Journel_Huijbregts_1978_book,Kitanidis_book}.  
It is known \cite{Fritz_Nowak_Neuweiler_2009_FFT_Kriging} that only the first column of the circulant matrix has to be stored. This reduces the computing cost from quadratic to log-linear \cite{Zimmerman_89} in $\bar  N$.
The key in the FFT-based Kriging is the fact that the multiplication of a circulant matrix and a vector is a discrete convolution which can be computed swiftly through FFT algorithm so that the quadratic computational  complexity is also reduced to a log-linear one \cite{Genton_2007EnMet_Separable_Covariance}.

If the measurements are given on a regular  equispaced grid, the first  Kriging task is solving a system also with a symmetric positive-definite Toeplitz matrix \cite{Fritz_Nowak_Neuweiler_2009_FFT_Kriging, Chan_Ng_1996}. Further development of methods handling
measurements that are on a subset of a finer regular grid have been made in \cite{Pegram_2004,Fritz_Nowak_Neuweiler_2009_FFT_Kriging}.

The work in \cite{NowakLitv2013} combined the power of FFT and the low-rank canonical tensor decomposition. It was assumed that the covariance matrix and the vector of interest (of size $\check N$) are available in a low-rank canonical tensor format which is a sum of $r$ Kronecker products of vectors of size $\check n$ each, with $\check n^d = \check N$. Separable covariance functions (e.g. Gaussian, separate exponential) can be decomposed exactly with $r=1$. For smooth non-separable covariance functions, a small $r$ value can usually give a good approximation. 

The canonical tensor representation can not only greatly reduce the memory storage size of the circulant matrix, but also speed up the Fourier transform since the $d$-dimensional FFT applied on the Kronecker product of matrices can be implemented by computing the 1-dimensional FFT on the first direction of each matrix. This reduces the complexity to $\mathcal{O}(d r \nc \log \nc)$. For $r \ll \nc$ this is a significant reduction from the complexity of FFT on the full tensor, which is $\mathcal{O}(d \nc^d \log \nc)$.

\subsection{Goals, approach and contributions} \label{sec:Goals}

However, converting a full tensor to a well approximating low-rank tensor format can be computationally formidable. Simply generating the full tensor itself might be beyond the memory capacity of a desktop computer. To make the low-rank FFT-based method practical, we need an efficient way to obtain a low-rank approximation directly from the multi-dimensional function that underlies the full tensor. 
It could be a challenging task though to approximate the first column of the Toeplitz (circulant) matrix in the canonical tensor format for $d \geq 3$. This is due to the fact that the class of rank-$k$ canonical tensors is a nonclosed set in the corresponding tensor product space (pp 91-92 in \cite{khorBook18}). 
The Tucker format tensor decomposition \cite{khor-survey-2011,hackbusch2012tensor,larskres-survey-2013} adopted in \cite{litvinenko2019tucker} could be too costly to use for problems with $d \geq 3$.

In this paper, we adopt an alternative tensor format, namely, the Tensor Train (TT) format \cite{osel-tt-2011,hackbusch2012tensor} (introduced in \refsec{sec:TT}) which can be obtained from a full tensor in a stable direct way by a sequence of singular value decompositions of auxiliary matrices, or, more importantly, it can be computed iteratively by the \emph{TT-cross} method \cite{ot-ttcross-2010} which has the complexity in the order of $\mathcal{O}(d r^3 \nee)$,
see \refsec{sec:TT-cross} for more details. 
Often this is the most time-consuming stage of Kriging operations.
Once the tensors are approximated in the TT format,
the FFT can be carried out with a modest $\mathcal{O}(d r^2 \nee \log \nee)$ complexity.
This makes the overall low-rank FFT-based Kriging practical for high dimensions.
We test the efficiency of the method in terms of computational time and memory usage in \refsec{sec:Performance-Tests}.

Thus, our paper is novel in three aspects: (i) we approximate the covariance matrix in the low-rank TT tensor format using only the given covariance function as a black box (this part was missing in \cite{NowakLitv2013}), (ii) we extend the methodology to Mat\'ern, exponential and spherical covariance functions (in addition to Gaussian functions), and (iii) we demonstrate that the low-rank approach enables high-dimensional Kriging. 

\subsection{Notation}

We denote vectors by bold lower-case letters (e.g., $\mathbf{c}$, $\mathbf{u}$, $\boldsymbol{\xi}$) and matrices by bold upper-case letters (e.g., $\mathbf{C}_{ss}$,  $\mathbf{M}$, $\mathbf{H}$). Letters decorated with an overbar represent the size of the tensor grid of estimates. Embedded matrices, vectors and their sizes are denoted by letters with a check accent  (e.g., $\Qsse$, $\qsse$, $\ne$, $\nie$). $\Fd$ stands for $d$-dimensional Fourier transform (FT), $\Fi$ for one-dimensional FT along the $i$-th dimension. $\iFd$ and  $\iFi$ are their inverse operators.

\section{Kriging and geostatistical optimal design\label{sec:Kriging_OD}}
Like in \cite{NowakLitv2013}, we work with the \emph{function estimate form  } \cite{Kitanidis_1996_2,Kitanidis_book} of Kriging (introduced in \refsec{sec:Simple_Kriging_dual_form}). We take simple Kriging  in which the estimates are assumed to have zero mean. 
\subsection{Mat\'{e}rn covariance}
\label{sec:Matern}
A low-rank approximation of the given function or a data set is a key component of the tasks formulated above.
Among of the many covariance models available, the Mat\'{e}rn family \cite{Matern_1986} 
is widely used in 
spatial statistics and
geostatistics. 

The Mat\'{e}rn covariance function is defined as
\begin{equation}
\label{eq:Matern}
C_{\nu,\ell}(r)=\frac{2^{1-\nu}}{\Gamma(\nu)}\left(\frac{\sqrt{2\nu}r}{\ell}
\right)^\nu K_\nu\left(\frac{\sqrt{2\nu}r}{\ell}\right).
\end{equation}
Here $r:=\Vert p_1-p_2\Vert $ is the distance between two points $p_1$ and $p_2$ in $\mathbb{R}^d$;
$\nu>0$ defines the smoothness. The larger is parameter $\nu$, the smoother is the random field.
The parameter $\ell>0$ is called the covariance length and measures how quickly the correlation of the random 
field decays with distance.  
${\cal K}_\nu$ denotes the modified Bessel function of order $\nu$. 
It is known that setting $\nu=1/2$ we obtain the exponential covariance model.
The value $\nu=\infty$ corresponds to a Gaussian covariance model.

In \cite{litvinenko2019tucker}, the authors provided the analytic $\sinc$-based 
proof of the existence of low-rank tensor approximations of Mat\'{e}rn functions.
They investigated numerically the behavior of
the Tucker and canonical ranks across a wide range of parameters specific to the family of Mat\'{e}rn kernels. 
It could be problematic to extend the results of this work to $d>3$, since one of the terms in the Tucker decomposition storage cost $\mathcal{O}(drn +r^d)$ is growing exponentially with $d$.
%

\subsection{Computational tasks in Kriging and optimal sampling design\label{sec:Simple_Kriging_dual_form}}
The computation of a simple Kriging process and optimal sample design involve mainly these   tasks:

\textbf{Task-1.} Let $\mathbf{y}$ denote a $N$-size  vector containing the sampled values,   $\mathbf{C}_{yy}$ denote the auto-covariance matrix.  If the measurements are not exact and the covariance matrix $\mathbf{R}$ of the random measurement error is available,  $\mathbf{R}$ is to be added to $\mathbf{C}_{yy}$. 
The first task is to solve  the below system for the Kriging weights $\boldsymbol{\xi}$:
\begin{equation}
 \mathbf{C}_{yy} \boldsymbol{\xi} = \mathbf{y} 
\label{eq:solve_weights}
\end{equation}

\textbf{Task-2.}
With the weights $\boldsymbol{\xi}$ we can obtain the Kriging estimates $\hat{\mathbf{s}}$ (sized $\Nee\times 1$ ) by a superposition of  columns of the cross-covariance matrices
 $\mathbf{C}_{sy}$ (sized $\Nee \times N$ ) weighted by $\boldsymbol{\xi}$, i.e. 
 the Kriging estimate $\hat{\mathbf{s}}$ is given by \cite{Kitanidis_book}:

\begin{equation}
\hat{\mathbf{s}}=\mathbf{C}_{sy}  \boldsymbol{\xi}\,.
\label{eq:estimate}
\end{equation}

\textbf{Task-3.} 
The variance $\hat{\bm{\sigma}}^2_{\mathbf{s}}$ of the estimates $\hat{\mathbf{s}}$ is to be obtained from the diagonal of the conditional covariance matrix $\mathbf{C}_{ss|y}$: 
\begin{eqnarray}
\hat{\boldsymbol{\sigma}}^2_{\mathbf{s}} = \diag(\mathbf{C}_{ss|y})
&=&\diag\left( \mathbf{C}_{ss}           - \mathbf{C}_{sy} \mathbf{C}_{yy}^{-1}\mathbf{C}_{ys} \right) \nonumber \\
&=&\diag\left( \mathbf{C}_{ss} \right) - \sum_{i=1}^N \left( \mathbf{C}_{sy}\boldsymbol\zeta_i\right)^{\circ 2}, 
\label{eq:estvar}
\end{eqnarray}
where $\boldsymbol\zeta_i$ is the $i$-th column of $ \mathbf{L}^{-T}$ with $\mathbf{L}$ the lower triangular Cholesky factor matrix of $\mathbf{C}_{yy}$, and the superscript $\circ 2$ denotes Hadamard square.


\textbf{Task-4.} The goal of geostatistical design is to optimize sampling patterns (or locations) for $\mathbf{y}$.
There two most common objective functions to be minimized, which are also called $A$- and $C$- criteria of geostatistical optimal design
\cite{Mueller_2007_Spatial_Data,Nowak_2010MG_Measures_Uncertainty, Cirpka_Nowak_2004_TravelTime}:
\begin{eqnarray}
\phi_A & = & \Nee^{-1}\trace\left[\mathbf{C}_{ss|y}\right] \nonumber \\
\phi_C & = & \mathbf{z}^{\top}\mathbf{C}_{ss|y}\mathbf{z}=\mathbf{z}^{\top}(\mathbf{C}_{ss}-\mathbf{C}_{sy}\mathbf{C}_{yy}^{-1}\mathbf{C}_{ys})\mathbf{z} \, ,
\label{eq:OD_criteria}
\end{eqnarray}
where $\mathbf{z}$ is a data vector  \cite{Nowak_2010MG_Measures_Uncertainty}.
%
\section{Interface from Kriging to FFT-based methods\label{sec:KrigingFFT}}
In this section we give a brief introduction to the basics of FFT-based Kriging \cite{Fritz_Nowak_Neuweiler_2009_FFT_Kriging}.
We assume that the measurement points are a subset of the estimate grid points.
The simplest version of Kriging is a direct \emph{injection}: the estimated values are set equal to the measurement values at the corresponding locations, and to zeros at all other points.
Equivalently, we say that we inject a (small) tensor of measurements into a (larger) tensor of estimations.

For the FFT-based Kriging we use a regular, equispaced grid which leads to a (block) Toeplitz covariance matrix that can be augmented to a circulant one \refsecp{Sub:Toeplitz}. 
An embedding operation augments the injected tensor to the size that is compatible with the circulant covariance matrix.  
The (pseudo-)inverse of embedding
is called extraction \refsecp{sub:sampling_injection}.

\subsection{Embedding Toeplitz covariance to circulant   matrices\label{Sub:Toeplitz}}

A Toeplitz matrix is constant along each descending diagonal (from left to right). A block Toeplitz matrix has identical sub-matrices in each descending diagonal block and each sub-matrix Toeplitz. If the covariance function is stationary and the estimates are made on a $d$-dimensional regular, equispaced grid, the covariance matrix $\mathbf{C}_{ss}$ is symmetric level-$d$ block Toeplitz  \cite{Matrices_Methods_and_Applications}. 
Since submatrices are repeating along diagonals the required storage could be reduced from $\mathcal{O}(\Nee^{2})$ to $\mathcal{O}(\Nee)$ elements \cite{Zimmerman_89,Kailath_Sayed_1995}.

A circulant matrix $\Qsse$ is a Toeplitz matrix that has its first column $\qsse$ periodic. This type of matrices come from covariance functions that are periodic in the domain.
A circulant matrix-vector product can be computed efficiently by FFT \cite{Computational_FFT}. The eigenvalues of $\Qsse$ can be computed as the Fourier transform of its first column $\qsse$
\cite[pp. 350-354]{Varga_54,Matrices_Methods_and_Applications}.
These properties lead us to the fast FFT-based kriging methods.

A Toeplitz  matrix $\mathbf{C}_{ss}$ can always be augmented to  a  
circulant matrix $\Qsse$. This process is called \emph{embedding}. Let $\qss$ be the first column of $\mathbf{C}_{ss}$.
Embedding is often done by appending the second through the last but one element of $\qss$ to the end of $\qss$ in reverse order, which makes a periodic vector $\qsse$. For the cases $d>1$, this augmentation has to be done recursively in every level for the $d$-level Toeplitz covariance matrix.  An equivalent way of doing this is to augment the domain (to be $2^d$ times larger) and extend the covariance function to be periodic on the domain, as illustrated in
\cite{Kozintsev_1999a,Nowak_al_2003}. In \cite{Newsam_Dietrich_94,Dietrich_Newsam_97_SIAM,Nowak_al_2003} the authors have addressed the issue of the  minimum embedding size. 

  
\subsection{Injection, embedding and extraction of data tensors \label{sub:sampling_injection}} 
      
Suppose we obtained the Kriging weights $\vek \xi$ for the measurements by solving \eqref{eq:solve_weights}. The injection of $\vek \xi$ means to insert it in a larger all-zero tensor that has the same size of the estimate tensor, i.e. the \emph{injected}  tensor has non-zero entries only at the measurement sites.

Suppose we have $N$ measurements indexed by $j =1, \cdots, N$, each  associated with a weight $\xi_j$ and a site index vector $\vek \alpha_j$, then the injection of $\vek \xi$ results in a tensor $\bar{\vek \xi} \in\mathbb{R}^{ \bar n_1 \times  \bar n_2 \times \cdots \times  \bar n_d}$  with entries:
\begin{equation} 
\bar{\vek \xi}(i_1, i_2, \cdots, i_d)=\left\{ \begin{array}{l}
\xi_j \quad \mbox{if } \vek i = \vek \alpha_j, \forall j \in [1,\cdots, N] \\
0\quad\mbox{otherwise}\end{array}\right.\,\,.\label{eq:injected_tensor}
\end{equation}
We denote the injection  operation by $\mathcal{H}:  \vek \xi  \rightarrow \bar{\vek \xi} $.

Embedding an injected  weight tensor enhances its mode size from $\nee$ to $\ne=2 \nee$ by padding zeros to the extra entries so that the tensor is of $2^d$ times the original size.  The embedded  weight tensor 
  $\check{\vek \xi} \in\mathbb{R}^{ \check n_1 \times  \check n_2 \times \cdots \times  \check n_d}$  has entries:
\begin{equation} 
\check{\vek \xi}(i_1, i_2, \cdots, i_d)=\left\{ \begin{array}{ll}
\bar{\vek \xi}(i_1, i_2, \cdots, i_d) &\quad \mbox{if }  i_\ell \leq \nee_\ell, \;  1\leq \ell \leq d \\
0 &\quad\mbox{otherwise}\end{array}\right.\,\,.\label{eq:embedded_tensor}
\end{equation}
We denote the embedding operation by $\mathcal{M}: \bar{\vek \xi} \rightarrow \check{\vek \xi} $. 

The  extraction is the inverse operation of embedding, we denoted it by $\mathcal{M}^{\dagger}$. By $\mathcal{M}^{\dagger}(\vek \eta)$ we take only the first half of $\vek \eta$ in every dimension, which results in a new tensor of only $\frac{1}{2^d}$ of the size of $\vek \eta$.

\subsection{Matrix-vector multiplication via FFT}\label{sub:Convolution-via-FFT}
With the circulant covariance matrix $\Qsse$ obtained as explained in \refsec{Sub:Toeplitz}, the Task-2 in \eqref{eq:estimate} becomes  a discrete convolution which can be computed by using FFT\cite{Computational_FFT}, this is written as (e.g., Fritz, Nowak and Neuweiler, \cite{Fritz_Nowak_Neuweiler_2009_FFT_Kriging}):
\begin{align}
   \mathbf{C}_{sy}\boldsymbol{\xi} =\mathbf{C}_{ss}\mathcal{H}(\vek \xi) &=\mathcal{M}^{\dagger}\Qsse\mathcal{M}(\mathcal{H}(\vek \xi)) \nonumber \\ 
   &=\mathcal{M}^{\dagger}\iFd \left (\Fd \left( \qsse\right)\circ \Fd \left(\check{ \vek \xi}\right)\right)\,\,.
\label{eq:superposition_FFT}
\end{align} 
where the operation $\mathcal{M}(\mathcal{H}( \cdot ))$ injects and embeds $\boldsymbol{\xi}$ into  $\check{\vek \xi}$. The $\Fd$ is evaluated by the Fast Fourier Transformation (FFT) \cite{FFTW}.
Without using tensor approximations the computational complexity for Kriging is reduced to
$\mathcal{O}\left(\Ne \log \Ne \right)$, and the storage size reduced to $\mathcal{O}\left(\Ne \right)$.

For the variance estimation (Task-3) in  \eqref{eq:estvar} the FFT method also applies. We first need to do a Cholesky decomposition   $\mathbf{C}_{yy}=\mathbf{L}\mathbf{L}^T$, and inject and embed each  column $\vek \zeta_i$ of $\mathbf{L}^{-T}$ to get the corresponding $\Check{\vek \zeta_i}$. Then \eqref{eq:estvar}  can be computed as 
\begin{eqnarray}
\hat{\boldsymbol{\sigma}}^2_{\mathbf{s}}   
=& \sigma^2_s \vek 1_{\bar N}  - \sum_{i=1}^N\left[ \mathcal{M}^{\dagger} \iFd \left (\Fd \left( \qsse\right)\circ \Fd \left(\check{ \vek \zeta_i}\right) \right)   \right]^{\circ 2},
\label{eq:estvar_FFT}
\end{eqnarray}
where $\sigma^2_s$ is the prior variance, $\vek 1_{\bar N}$ is a $\bar N$-length vector of all ones.

\section{FFT-based Kriging accelerated by low-rank tensor decomposition \label{sec:low_rank_FFT_Kriging}}
In addition to the efficient FFT-based method enabled by the Teoplitz structure of covariance matrices, the Kriging process can be further sped up by low-rank representations of the embedded covariance matrices. Since the covariance functions are usually smooth, large covariance matrices could be well approximated by a low-rank tensor format.
A  literature  survey  of  low-rank  tensor approximation techniques is available in \cite{khor-survey-2011,larskres-survey-2013}.

In this section, we approximate the first column of the circulant covariance matrix in tensor train (TT) format and then rewrite   \refeq{eq:superposition_FFT} also in the TT format. We start with a brief reviewing of the TT technique.
\subsection{TT decomposition}
\label{sec:TT}
We assume that the data vectors ($\mathbf{c}$, $\boldsymbol{\xi}$, etc.) can be associated to a function discretised on a structured grid in $d$ dimensions, for example, if $u(x,y,z)$ is sampled on a Cartesian 3-dimensional grid,
\begin{equation}
\boldsymbol{\xi} = \left\{\xi(x_{i_1},y_{i_2},z_{i_3})\right\}_{i_1,i_2,i_3=1}^{n_1,n_2,n_3}.
\end{equation}
Then we can enumerate the entries of the vector via sub-indices $i_1,i_2,\ldots,i_d$, thereby seeing it as a \emph{tensor} with elements
$
\boldsymbol{\xi}(i_1,\ldots,i_d).
$
We approximate such tensors, and, consequently, associated data vectors, in the Tensor Train (TT) decomposition \cite{osel-tt-2011},
\begin{equation}
 \boldsymbol{\xi}(i_1,i_2,\ldots,i_d) \approx \boldsymbol{\tilde\xi}(i_1,i_2,\ldots,i_d) :=  \sum_{\alpha_0,\ldots,\alpha_d=1}^{r_0,\ldots,r_d}\xi^{(1)}_{\alpha_0,\alpha_1}(i_1)\xi^{(2)}_{\alpha_1,\alpha_2}(i_2) \cdots \xi^{(d)}_{\alpha_{d-1},\alpha_d}(i_d).
 \label{eq:tt}
\end{equation}
Here $\xi^{(k)}$, $k=1,\ldots,d$, are called \emph{TT blocks}.
Each TT block $\xi^{(k)}$ is a three-dimensional tensor of size $r_{k-1} \times n_k \times r_k$, $r_0=r_d=1$.
The efficiency of this representation relies on the \emph{TT ranks} $r_0,\ldots,r_d$ being bounded by a moderate constant $r$.
For simplicity we can also introduce an upper bound of the univariate grid sizes $n_k \le n$.
Then we can notice that the TT format \eqref{eq:tt} contains at most $dnr^2$ elements.
This is much smaller than the number of entries in the original tensor which grows exponentially in $d$.
Using Kronecker products, one can rewrite \eqref{eq:tt} as follows,
$$
\boldsymbol{\tilde \xi} =  \sum_{\alpha_0,\ldots,\alpha_d=1}^{r_0,\ldots,r_d}\xi^{(1)}_{\alpha_0,\alpha_1} \otimes \xi^{(2)}_{\alpha_1,\alpha_2} \otimes \cdots  \otimes \xi^{(d)}_{\alpha_{d-1},\alpha_d},
$$
i.e. we see each TT block as a set of vectors of length $n_k$.

Of course, one can think of any other scheme of sampling a function, e.g. at random points, but the TT decomposition requires independence of sub-indices $i_1,\ldots,i_d$,
and therefore the Cartesian product discretisation.
The rationale behind using this, on the first glance excessive, scheme,
is the fast convergence of the approximation error $\varepsilon$ with the TT ranks.
If $\xi(x,y,z)$ is analytic, the TT ranks often depend logarithmically on $\varepsilon$ \cite{tee-tensor-2003,khor-rstruct-2006,uschmajew-approx-rate-2013}.
Combining the TT approximation with collocation on the Chebyshev grid, which allows to take $n = \mathcal{O}(|\log \varepsilon|)$ for analytic functions,
one arrives at $\mathcal{O}(d|\log\varepsilon|^{3})$ overall cost of interpolation or integration using the TT format.
This can be significantly cheaper than the $\mathcal{O}(\varepsilon^{-2})$ cost of Monte Carlo quadrature or Radial Basis function interpolation.
Moreover, TT ranks depend usually very mildly on the particular univariate  discretisation scheme, provided that it can resolve the function.
We can use any univariate grid in each variable instead of the Chebyshev rule.
For example, a uniform grid yields Toeplitz or circulant covariance matrices, which are amenable to fast FFT-based multiplication/diagonalisation.

However, it is difficult to obtain sharp bounds for the TT ranks theoretically.
Therefore, we resort to robust numerical algorithms to compute a TT approximation of given data.

\subsection{TT-cross approximation} \label{sec:TT-cross}
A full tensor can be compressed into a TT format quasi-optimally for the desired tolerance via the truncated singular value decomposition (SVD)~\cite{osel-tt-2011}. 
However, the full tensor might even be impossible to store. 
In this section we recall the practical TT-cross method \cite{ot-ttcross-2010} that computes the representation~\eqref{eq:tt} using \emph{only a few} entries from $\boldsymbol{\xi}$.
It is based on the skeleton decomposition of a matrix~\cite{TyrtyshACA}, which represents an $n\times m$ matrix $A$ of rank $r$ as the \emph{cross} (in Matlab-like notation)
\begin{equation}
 A=A(:,\mathcal{J})A(\mathcal{I},\mathcal{J})^{-1}A(\mathcal{I},:)
 \label{eq:cross}
\end{equation}
of $r$ columns and rows, where $\mathcal{I}$ and $\mathcal{J}$ are two index sets of cardinality $r$ such that $A(\mathcal{I},\mathcal{J})$ (the intersection matrix) is invertible. 
If $r\ll n,m$, the right-hand side requires only $(n+m-r)r\ll nm$ elements of the original matrix.

In order to describe the TT-cross method, we introduce the so-called
\emph{unfolding} matrices
$\Xi_k=[\boldsymbol{\xi}(i_1,\ldots,i_k;i_{k+1},\ldots,i_d)]$, that have the first $k$ indices grouped together to index rows, and the remaining indices grouped to index columns.
Let us now consider $\Xi_1$ and apply the idea of the matrix cross \eqref{eq:cross}.
Assume that there exists a set of $r_1$ index tuples, $\mathcal{I}_{>1} = \{i_2^{\alpha_1},\ldots,i_d^{\alpha_1}\}_{\alpha_1=1}^{r_1}$,
such that the $\mathcal{I}_{>1}$-``columns'' of the original tensor $\boldsymbol{\xi}(:,\mathcal{I}_{>1})$ form a ``good'' basis for all columns of $\Xi_1$.
The reduction~\eqref{eq:cross} may be formed for $r_1$ rows at positions $\mathcal{I}_{<2} = \{i_1^{\alpha_1}\}_{\alpha_1=1}^{r_1}$, which are now \emph{optimized} by choosing the $r_1 \times r_1$ submatrix $\boldsymbol{\xi}(\mathcal{I}_{<2},\mathcal{I}_{>1})$ such that its \emph{volume} (modulus of determinant) is maximal.
This can be done by the \emph{maxvol} algorithm \cite{gostz-maxvol-2010} in $\mathcal{O}(nr_1^2)$ operations.
Now we construct the first TT block $\xi^{(1)}$ as the $n \times r_1$ matrix $\boldsymbol{\xi}(:,\mathcal{I}_{>1})\boldsymbol{\xi}(\mathcal{I}_{<2},\mathcal{I}_{>1})^{-1}$.
In a practical algorithm, the inversion is performed via the QR-decomposition for numerical stability.
Next, we reduce the tensor onto $\mathcal{I}_{<2}$ in the first variable, and apply TT-cross inductively to
$[\Xi_{>1}({\alpha_1}, i_2,\ldots,i_d)] = [\boldsymbol{\xi}(i_1^{\alpha_1}, i_2,\ldots,i_d)]$.

In the $k$-th step, assume that we are given the reduction $\Xi_{>k-1}(\alpha_{k-1},i_k,\ldots,i_d)$,
 a ``left'' index set $\mathcal{I}_{<k} = \{i_1^{\alpha_{k-1}},\ldots,i_{k-1}^{\alpha_{k-1}}\}_{\alpha_{k-1}=1}^{r_{k-1}}$,
and a ``right'' set $\mathcal{I}_{>k} = \{i_{k+1}^{\alpha_k},\ldots,i_d^{\alpha_k}\}_{\alpha_k=1}^{r_k}$.
The $r_{k-1}n \times r_k$ reduced unfolding matrix $[\Xi_{>k-1}({\alpha_{k-1}}, i_k;~\mathcal{I}_{>k})]$ is again feasible for the \emph{maxvol} algorithm, which produces a set of row positions $\ell_{k} = \{\alpha_{k-1}^{\alpha_k}, i_k^{\alpha_k}\}_{\alpha_k=1}^{r_k}$.
The next left set $\mathcal{I}_{<k+1}$ is constructed from $\ell_k$ by replacing $\alpha_{k-1}$ with the corresponding indexes $i_1^{\alpha_{k-1}},\ldots,i_{k-1}^{\alpha_{k-1}}$ from $\mathcal{I}_{<k}$.
Continuing this process until the last variable, where we just copy $\xi^{(d)} = \Xi_{>d-1}$, we complete the induction.

This process can be also organized in a form of a binary tree, which gives rise to the so-called hierarchical Tucker cross algorithm \cite{bg-htcross-2015}.
In total, we need $\mathcal{O}(dnr^2)$ evaluations of $\boldsymbol{\xi}$ and $\mathcal{O}(dnr^3)$ additional operations in computations of the maximum volume matrices.

The TT-cross method requires some starting index sets $\mathcal{I}_{>k}$.
Without any prior knowledge, it seems reasonable to initialize $\mathcal{I}_{>k}$ with independent realizations of
any easy to sample reference distribution (e.g. uniform or Gaussian).
If the target tensor $\boldsymbol{\xi}$ admits an \emph{exact} TT decomposition with TT ranks not greater than $r_1,\ldots,r_{d-1}$,
and all unfolding matrices have ranks not smaller than the TT ranks of $\boldsymbol{\xi}$,
the cross iteration outlined above reconstructs $\boldsymbol{\xi}$ \emph{exactly} \cite{ot-ttcross-2010}.
However, practical tensors can usually only be \emph{approximated} by a TT decomposition with low ranks.
Nevertheless a slight \emph{overestimation} of the ranks can deliver a good approximation, if a tensor was produced from a regular enough function \cite{bg-htcross-2015,ds-alscross-2017pre}.

However, it might be necessary to refine the sets $\mathcal{I}_{<k},\mathcal{I}_{>k}$ by conducting \emph{several} TT cross iterations, going back and forth over the TT blocks and optimizing the sets by the maxvol algorithm.
For example, after computing $\xi^{(d)} = \Xi_{>d-1}$, we ``reverse'' the algorithm and apply the maxvol method to the \emph{columns} of a $r_{d-1} \times n$ matrix $\xi^{(d)}$. This gives a \emph{refined} set of points $\mathcal{I}_{>d-1} = \{i_d^{\alpha_{d-1}}\}$.
The recursion continues from $k=d$ to $k=1$, optimizing the right sets $\mathcal{I}_{>k}$, while taking the left sets $\mathcal{I}_{<k}$ from the previous (forward) iteration.
After several iterations, both $\mathcal{I}_{<k}$ and $\mathcal{I}_{>k}$ can be optimized to the particular target function,
even if the starting sets were inaccurate.

This adaptation of points can be combined with the \emph{adaptation of ranks}.
If the initial ranks $r_1,\ldots,r_{d-1}$ were too large, they can be reduced to quasi-optimal values for the desired accuracy via SVD.
However, we can also \emph{increase} the ranks by computing the unfolding matrix $\left[\mathbf{u}(\mathcal{I}_{<k}, i_k;~i_{k+1}^{\alpha_k},\ldots,i_{d}^{\alpha_k})\right]$ on an \emph{enriched} index set: we take $\{i_{k+1}^{\alpha_k},\ldots,i_d^{\alpha_k}\}$ from $\mathcal{I}_{>k}$ for $\alpha_k=1,\ldots,r_k$, and also from an \emph{auxiliary} set $\mathcal{I}_{>k}^{aux}$ for $\alpha_k=r_k+1,\ldots,r_k+\rho$.
This increases the $k$-th TT rank from $r_k$ to $r_k+\rho$.
The auxiliary set can be chosen at random~\cite{Os-mvk2-2011} or using a surrogate for the error~\cite{ds-amen-2014}. 
The pseudocode of the entire TT cross method is listed in Algorithm \ref{alg:cross},
where we let $\mathcal{I}_{<1} = \mathcal{I}_{>d} = \emptyset$ for uniformity.
\begin{algorithm}[t]
\caption{TT cross algorithm with rank adaptation.}
\label{alg:cross}
\begin{algorithmic}[1]
 \Require Initial index sets $\mathcal{I}_{>k}$, rank increasing parameter $\rho\ge 0$, stopping tolerance $\delta>0$ and/or maximum number of iterations $\mathrm{iter}_{\max}$.
 \Ensure TT blocks of an approximation \eqref{eq:tt} to $\boldsymbol{\xi}$.
 \While{$\mathrm{iter}<\mathrm{iter}_{\max}$ and $\|\boldsymbol{\tilde\xi}_{\mbox{iter}}-\boldsymbol{\tilde\xi}_{\mbox{iter}-1}\|>\delta \|\boldsymbol{\tilde\xi}_{\mbox{iter}}\|$}
    \For{$k=1,2,\ldots,d$} \Comment{Forward iteration}
      \State (Optionally) prepare an auxiliary enrichment set $\mathcal{I}_{>k}^{aux}$.
      \State Compute the $r_{k-1}n \times r_k$ unfolding matrix $\boldsymbol{\xi}(\mathcal{I}_{<k},i_k;~\mathcal{I}_{>k})$.
       \State Compute $\mathcal{I}_{<k+1}$ by the \emph{maxvol} algorithm and (optionally) truncate.
    \EndFor
    \For{$k=d,d-1,\ldots,1$} \Comment{Backward iteration}
      \State (Optionally) prepare an auxiliary enrichment set $\mathcal{I}_{<k}^{aux}$.
      \State Compute the $r_{k-1} \times nr_k$ unfolding matrix $\boldsymbol{\xi}(\mathcal{I}_{<k}~;i_k,\mathcal{I}_{>k})$.
    \State Compute $\mathcal{I}_{>k-1}$ by the \emph{maxvol} algorithm and (optionally) truncate.
    \EndFor
 \EndWhile
\end{algorithmic} 
\end{algorithm} 
Empowered with the enrichment scheme, we are not limited to just truncating ranks from above.
Instead, we can start with a low-rank initial guess and increase the ranks until the desired accuracy is met.

\subsection{TT representation of general and structured matrices}
Let us now consider how the TT format \eqref{eq:tt} can be generalised to matrices $\mathbf{C} \in \mathbb{R}^{n^d \times n^d}$, such as the $\mathbf{C}_{ss}$ matrix from \eqref{eq:estvar}. 
Using sub-indices $i_1,\ldots,i_d$, we can think of a matrix as a $2d$-dimensional tensor with elements $\mathbf{C}(i_1,\ldots,i_d;~j_1,\ldots,j_d)$.
However, most matrices in our applications have full ranks, and a straightforward $2d$-dimensional TT decomposition would be inefficient.
Instead, we consider a permuted, or \emph{matrix} TT decomposition~\cite{osel-tt-2011}:
\begin{equation}
  \mathbf{C}(i_1,\ldots,i_d;~j_1,\ldots,j_d) = \sum_{\beta_0,\ldots,\beta_d=1}^{R_0,\ldots,R_d} C^{(1)}_{\beta_0,\beta_1}(i_1,j_1) C^{(2)}_{\beta_1,\beta_2}(i_2,j_2) \cdots C^{(d)}_{\beta_{d-1},\beta_d}(i_d,j_d),
 \label{eq:ttm}
\end{equation}
or in the Kronecker form,
\begin{equation}
  \mathbf{C} = \sum_{\beta_0,\ldots,\beta_d=1}^{R_0,\ldots,R_d} C^{(1)}_{\beta_0,\beta_1} \otimes C^{(2)}_{\beta_1,\beta_2} \otimes \cdots \otimes C^{(d)}_{\beta_{d-1},\beta_d}.
 \label{eq:ttm-kron}
\end{equation}

The identity matrix can be trivially represented in matrix TT format $I_{n^d} = I_n \otimes \cdots \otimes I_d$ with $R_0=\cdots =R_d=1$.
Furthermore, we can quickly assemble block Toeplitz and circulant matrices if their first column/row is given in the TT format \cite{khkaz-conv-2013}.
Let us introduce the operation $\mathcal{T}: \mathbb{R}^{2n} \rightarrow \mathbb{R}^{n \times n}$ which assembles a Toeplitz matrix from a vector of its first column and row stacked together, and the operation $\mathcal{C}: \mathbb{R}^{n} \rightarrow \mathbb{R}^{n \times n}$ which assembles a circulant matrix from its first column.
Assume that a vector $\mathbf{c}$ of size $(2n)^d$ or a vector $\mathbf{\check c}$ of size $n^d$ are given in the TT format \eqref{eq:tt},
\begin{equation}
\mathbf{c} = \sum_{\alpha_0,\ldots,\alpha_d=1}^{r_0,\ldots,r_d} c^{(1)}_{\alpha_0,\alpha_1} \otimes \cdots \otimes c^{(d)}_{\alpha_{d-1},\alpha_d}, \quad
\mathbf{\check c} = \sum_{\alpha_0,\ldots,\alpha_d=1}^{r_0,\ldots,r_d} \check c^{(1)}_{\alpha_0,\alpha_1} \otimes \cdots \otimes \check c^{(d)}_{\alpha_{d-1},\alpha_d}
\label{eq:ttc}
\end{equation}
Then the block Toeplitz or circulant matrix, respectively
$$
\mathbf{C} = \left(\bigotimes_{k=1}^{d} \mathcal{T}\right) \mathbf{c}, \qquad \mathbf{\check C} = \left(\bigotimes_{k=1}^{d} \mathcal{C}\right) \mathbf{\check c},
$$
can be written in the matrix TT formats \eqref{eq:ttm} with the same TT ranks,
$$
\mathbf{C} = \sum_{\alpha_0,\ldots,\alpha_d=1}^{r_0,\ldots,r_d} \left(\mathcal{T} c^{(1)}_{\alpha_0,\alpha_1}\right) \otimes \cdots \otimes \left(\mathcal{T} c^{(d)}_{\alpha_{d-1},\alpha_d}\right), \qquad 
\mathbf{\check C} = \sum_{\alpha_0,\ldots,\alpha_d=1}^{r_0,\ldots,r_d} \left(\mathcal{C} \check c^{(1)}_{\alpha_0,\alpha_1}\right) \otimes \cdots \otimes \left(\mathcal{C} \check c^{(d)}_{\alpha_{d-1},\alpha_d}\right).
$$
Similarly we can apply the multivariate Fourier transform without changing TT ranks:
\begin{equation}
\left(\bigotimes_{k=1}^{d} \mathcal{F} \right) \mathbf{c} = \sum_{\alpha_0,\ldots,\alpha_d=1}^{r_0,\ldots,r_d} \left(\mathcal{F} c^{(1)}_{\alpha_0,\alpha_1}\right) \otimes \cdots \otimes \left(\mathcal{F} c^{(d)}_{\alpha_{d-1},\alpha_d}\right),
\label{eq:tt-fft}
\end{equation}
where $\mathcal{F}: \mathbb{R}^{n} \rightarrow \mathbb{R}^{n}$ is the univariate FFT.
This reduces the complexity of FFT from $\mathcal{O}(N \log N) = \mathcal{O}(dn^d \log n)$ to $\mathcal{O}(dr^2 n \log n)$.

In general, the TT format allows to represent the product of any matrix given in \eqref{eq:ttm} and a compatible vector given in \eqref{eq:tt} in another TT format~\cite{osel-tt-2011} with multiplied ranks,
\begin{equation}
\mathbf{C}\boldsymbol{\xi} = \sum_{\gamma_0,\ldots,\gamma_d=1}^{(r_0R_0),\ldots,(r_dR_d)} \left(C^{(1)}_{\beta_0,\beta_1} \xi^{(1)}_{\alpha_0,\alpha_1}\right)_{\gamma_0,\gamma_1} \otimes \cdots \otimes \left(C^{(d)}_{\beta_{d-1},\beta_d} \xi^{(d)}_{\alpha_{d-1},\alpha_d}\right)_{\gamma_{d-1},\gamma_d},
\label{eq:ttmatvec}
\end{equation}
where $\gamma_k = \alpha_k + (\beta_k-1) r_k$, $k=0,\ldots,d$.

\subsection{Kriging operations in TT format}

To rewrite the Kriging estimation  \eqref{eq:superposition_FFT} in low rank format, we first find a TT approximation \eqref{eq:ttc} of $\mathbf{c}$
by using the TT-cross algorithm introduced in \refsec{sec:TT-cross}.
With the rest of the operations we can proceed in two ways.

\subsubsection{Small number of scattered samples}
If we assume $N$ to be small, the Task-1 of computing  Kriging weights, $C_{yy}\boldsymbol{\xi} = \mathbf{y}$, can be computed directly at low cost.
Now we inject the scattered values into a TT tensor of desired size as introduced in \eqref{eq:injected_tensor}.
Suppose $\boldsymbol{\ell}_j \in \mathbb{N}^{d}$ is the position of the $j$th sample, $j=1,\ldots,N,$ we can define 
$$
\mathcal{H}_j = \bigotimes_{k=1}^{d} \mathbf{e}_{j}^{(k)}, \quad \mbox{where} \quad \mathbf{e}_j^{(k)}(i_k) = \left\{\begin{array}{ll}1, & i_k=\ell_j(k) \\ 0, & \mbox{otherwise,} \end{array}\right.
$$
i.e. the injection operation \eqref{eq:injected_tensor} per sample.
Now the injected tensor is written in the CP format as
\begin{equation}
\boldsymbol{\bar \xi} = \sum_{j=1}^{N} \xi_j \mathcal{H}_j,
\label{eq:inject-tt}
\end{equation}
which can be converted to TT format directly by the formula in \cite[pp. 380]{Hackbusch_book} or using the Alternating Least Squares (ALS) \cite{holtz-ALS-DMRG-2012} approximation.

Similarly, we can use the direct truncation or the ALS method for summing columns of $\mathbf{C}_{sy}$ with the weights $\boldsymbol{\zeta}_i$ in \eqref{eq:estvar}, as well as the summation of different vectors $(\mathbf{C}_{sy} \boldsymbol{\zeta}_i)^{\circ 2}$.

Embedding operation \eqref{eq:embedded_tensor} is simpler and more efficient:
we just need to pad every TT block with zeros.
Assuming we are given a vector $\boldsymbol{\xi}$ in the form \eqref{eq:tt}, we construct the following new TT blocks of a vector $\boldsymbol{\check \xi}$:
\begin{equation}
\check\xi^{(k)}_{\alpha_{k-1},\alpha_{k}}(i_k) = \left\{\begin{array}{ll}\xi^{(k)}_{\alpha_{k-1},\alpha_k}(i_k), & i_k = 1,\ldots,\bar n_k, \\ 0, & i_k = \bar n_k+1, \ldots, n_k,\end{array}\right. \qquad k=1,\ldots,d.
\label{eq:embed_tt}
\end{equation}
Similarly, Extraction operation is performed by truncating the range of $i_k$ in each TT block from $n_k$ back to $\bar n_k$.
Most importantly, embedding and extraction can be performed very efficiently without changing the TT ranks, similarly to FFT \eqref{eq:tt-fft}.

Finally, we need to compute the Hadamard products of TT tensors, e.g. $\mathcal{F}^{[d]}(\mathbf{\check c}) \circ \mathcal{F}^{[d]}(\boldsymbol{\check \xi})$ in \eqref{eq:superposition_FFT}.
The Hadamard product
can be constructed exactly via \eqref{eq:ttmatvec} by noticing that \begin{equation*}
\mathbf{s} :=  \mathbf{c} \circ \boldsymbol{\xi} = \mathbf{C}\boldsymbol{\xi},\quad \text{for}\quad \mathbf{C}=\diag(\mathbf{c}),
\end{equation*} 
or approximately by applying the TT-Cross algorithm to a tensor given elementwise by the formula $
\mathbf{s}(i_1,\ldots,i_d) = \mathbf{c}(i_1,\ldots,i_d) \boldsymbol{\xi}(i_1,\ldots,i_d)$.
The direct multiplication requires $\mathcal{O}(dnR^2 r^2)$ operations, and the truncation afterwards has an even higher cost $\mathcal{O}(dnR^3 r^3)$.
In contrast, the TT-Cross approach needs computing $\mathcal{O}(dnr^2)$ samples of the target tensor $\mathbf{s}$, which means taking samples of the TT decompositions for $\mathbf{c}$ and $\boldsymbol{\xi}$ and multiplying them.
Sampling another TT tensor requires in total $\mathcal{O}(dnR^2 r)$ operations, which, assuming that the ranks are comparable, $R \sim r$, results in a total of $\mathcal{O}(dnr^3)$ operations in the TT-Cross computation of Hadamard products, which is thus preferred in this paper.

For geostatistical optimal design (Task-4)   we need to compute the trace of $\mathbf{C}_{ss|y}$. Since in the Task-3 we obtain already the diagonal of $\mathbf{C}_{ss|y}$ in the TT format, the trace can be evaluated swiftly by computing a dot product with the all-ones tensor.

\subsubsection{Large number of structured samples} \label{sec:largeN}
When $N$ is large, the summation \eqref{eq:inject-tt} can be a difficult operation in the TT format, potentially leading also to the TT ranks being in the order of $N$.
However, a large number of samples usually means that these samples are distributed fairly uniformly in the domain of interest.
In this case, we switch to the TT computations even before Task-1 in equation  \eqref{eq:solve_weights}.
First, we interpolate the given samples onto a uniform Cartesian grid with the mesh interval being in the order of the average distance between the original samples.
In the remaining operations, we assume that $\mathbf{y}$ is structured in this way, i.e. it can be seen as a tensor $\mathbf{y}(i_1,\ldots,i_d)$, $i_k=1,\ldots,\bar m_k$, $k=1,\ldots,d$.
Thus, we can approximate $\mathbf{y}$ in the TT format.

The solution for weights \eqref{eq:solve_weights}
becomes a rather difficult operation for a large $N$.
However, given the TT decompositions for $\mathbf{y}$ and $\mathbf{C}_{yy}$, the linear system can be solved more efficiently by employing ALS and similar tensor algorithms \cite{holtz-ALS-DMRG-2012,ds-amen-2014}.
Similarly, we can compute $\mathbf{C}_{yy}^{-1} \mathbf{C}_{ys}$ for \eqref{eq:estvar} by treating $\mathbf{C}_{ys}$ as the right hand side, and expanding $\mathbf{C}_{yy}$ accordingly.

If we interpolate $\mathbf{y}$ onto a periodic uniform Cartesian grid, the matrix $\mathbf{C}_{yy}$ becomes circulant, similarly to $\mathbf{\check C}$.
In this case we can approximate only its first column in the TT format, perform the Fourier transform to obtain the eigenvalues,
and apply again the TT-Cross method to approximate the pointwise division $\mathcal{F}^{[d]}(\mathbf{y})(i_1,\ldots,i_d)/\mathcal{F}^{[d]}(\mathbf{c})(i_1,\ldots,i_d)$.

\section{Numerical tests\label{sec:Performance-Tests}}


We used the Matlab package \emph{TT-Toolbox} ( \url{https://github.com/oseledets/TT-Toolbox}) for Tensor Train algorithms.
The codes used for numerical experiments are available at \url{https://github.com/dolgov/TT-FFT-COV}.
All computations are done on a MacBook Pro produced in 2013, equipped with 16GB RAM and an 2.7 GHz Intel Core i7 CPU.


We consider three test cases: 1) a 2-dimensional problem with $N=\prod_{i=1}^2 n_i=600^2$ (it is easy to visualize); 2) a 3-dimensional problem with $N=10^{15}$ and 3) 10-dimensional problem with $N=\prod_{i=1}^{10} n_i=100^{10}$. One of these parameters could be, for example, time.
%
The daily soil moisture data set, used below, is taken from \cite{Huang2016,LitvGenton2019,LitvGitHubMoisture}, where only one replicate, sampled at $N$ locations, is used.
\subsection{Kriging of daily moisture data}
\label{sec:moisture}
Numerical models play important role in climate studies.
These numerical models are complicated and high-dimensional, including such variables as pressure,
temperature, speed, and direction of the wind, level of precipitation, humidity, and moisture.
Many parameters are uncertain or even unknown.
Accurate modeling of soil moisture finds applications in the agriculture, weather prediction, early warnings of flood and in some others. Since the underlined geographical areas are usually large and high spatial resolutions are required, the involved data sets are huge. This could make the computational process in dense matrix format unfeasible or very expensive. By involving efficient low-rank tensor calculus, we can increase the spatial and time resolution and consider more parameters.  
It is clear that utilization of the rank $k$ tensor approximation introduces an additional numerical error in quantities of interest (QoIs). By increasing tensor ranks we reduce this approximation error.

\begin{figure}[h]
\center
\includegraphics[width=0.99\textwidth]{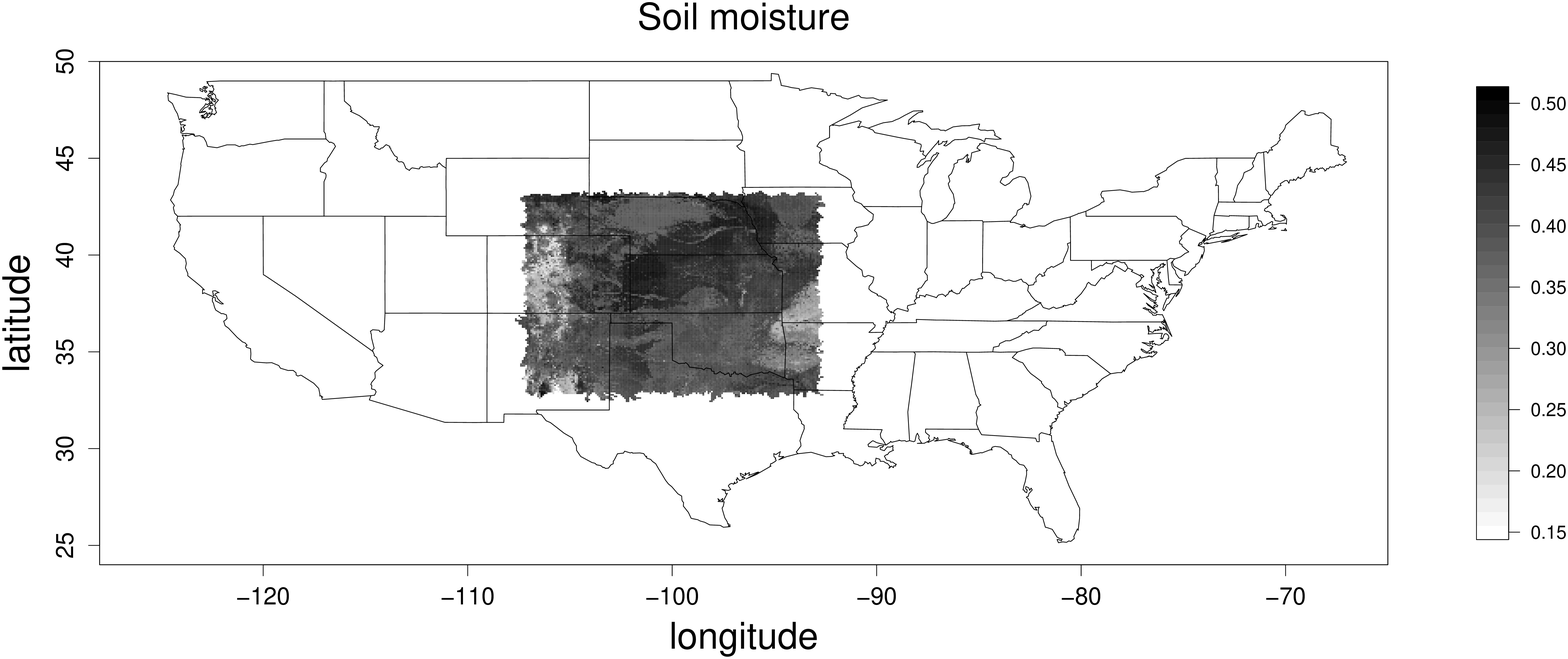}
\caption{The area where the daily soil moisture data were measured, Mississippi River basin, U.S.A.}
\label{fig:moistschema}
\end{figure}

We consider high-resolution soil moisture data from January 1, 2014, measured in the topsoil layer of the Mississippi River basin, U.S.A (Fig.~\ref{fig:moistschema}). 

\begin{figure}[h]
\center
\includegraphics[width=0.32\textwidth]{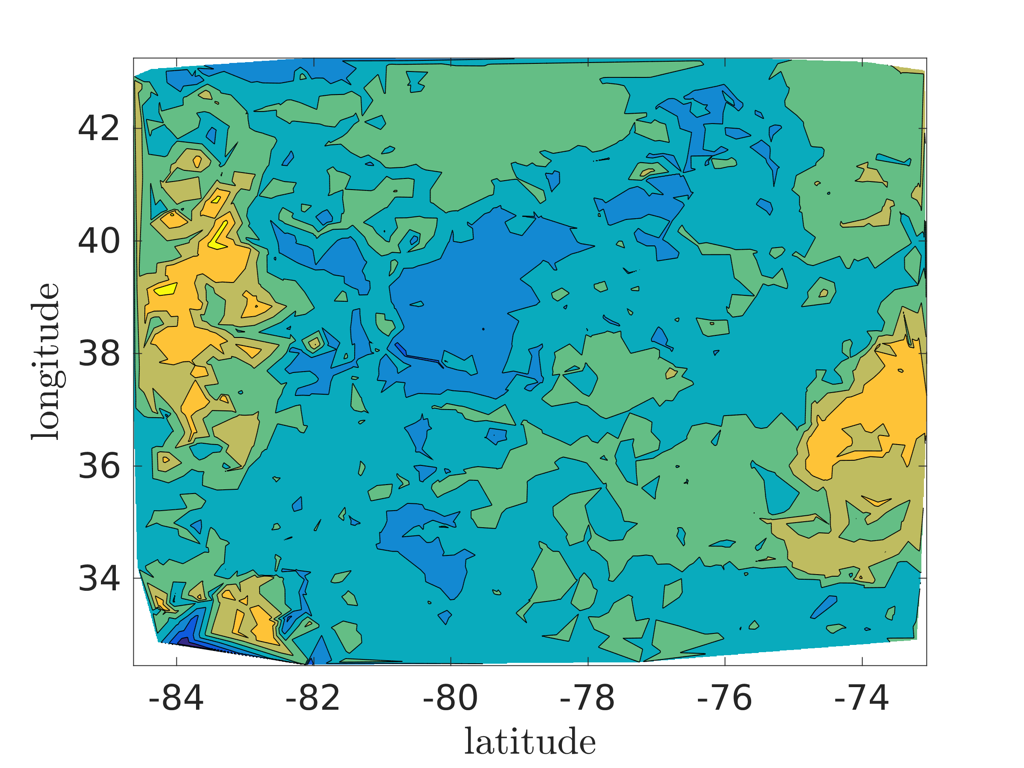}
\includegraphics[width=0.32\textwidth]{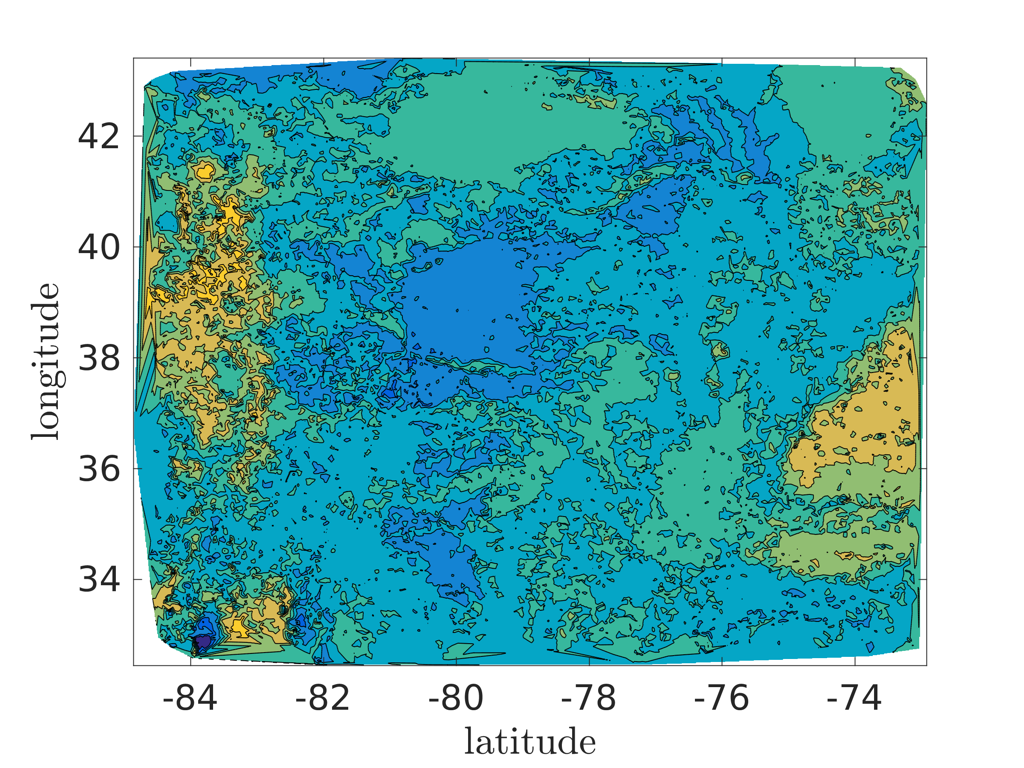}
\includegraphics[width=0.32\textwidth]{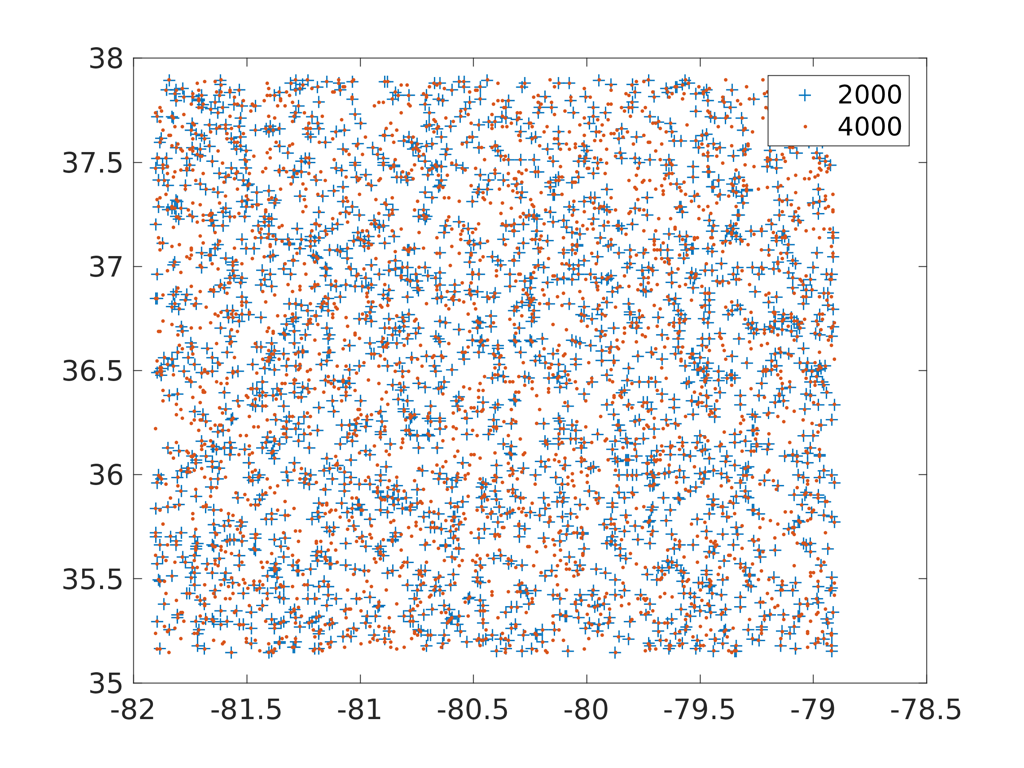}
\caption{Daily moisture data. Interpolated from (left) 2000 and (center) 40000 measurement points. (right) Two sets of sampling points, 2000 and 4000.}
\label{fig:moist}
\end{figure}

Figure~\ref{fig:moist} shows an example of daily moisture data. On the left picture we used 2000 points $(x,y,v)_{i=1}^N$, $N=2000$ for interpolation, and on the right 4000 points. The third picture shows two set of locations: one with 2000 points, marked with the blue symbol $+$ and with 4000 points, marked with red dot. 

The spatial resolution is 0.0083 degrees, and the distance of one-degree difference in this region is approximately 87.5 km. 
The grid consists of $1830 \times 1329 = 2{.}432{.}070$ locations with $2{.}000{.}000$ observations and $432{.} 070$ missing values. 
Therefore, the available spatial data are not on a regular grid. 

\begin{figure}[h]
\center
\includegraphics[width=0.34\textwidth]{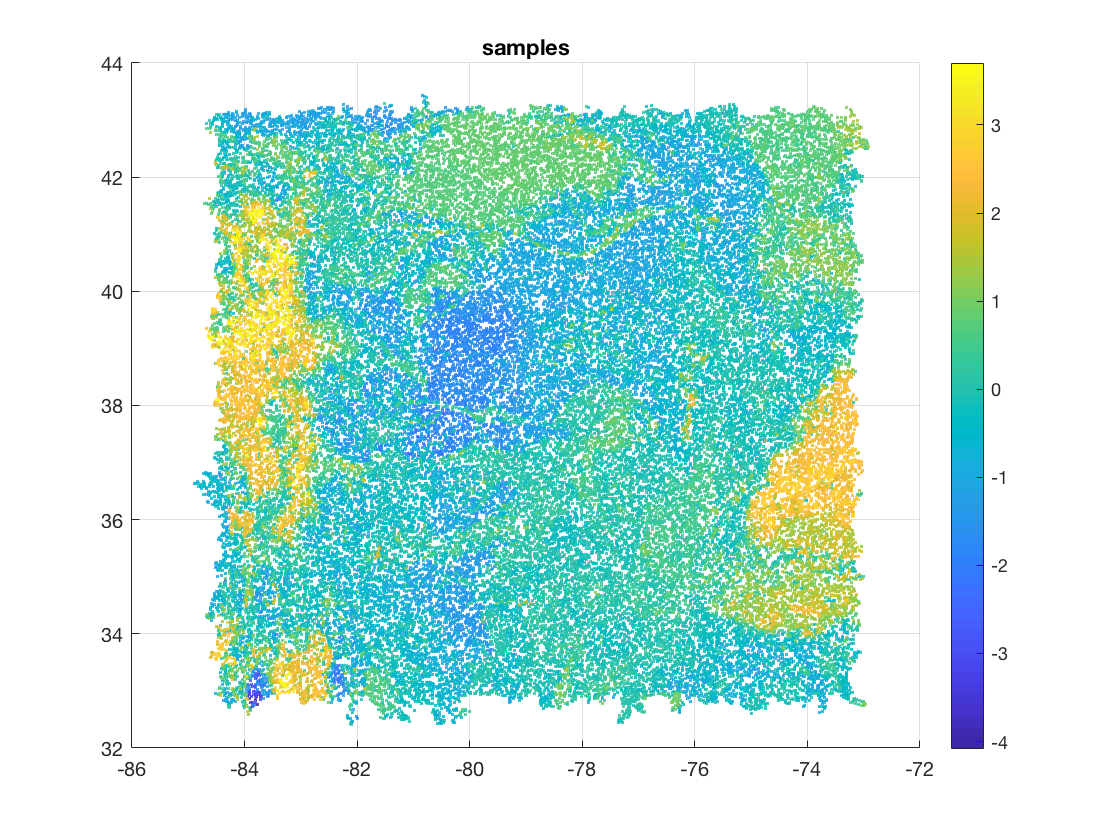}
\includegraphics[width=0.32\textwidth]{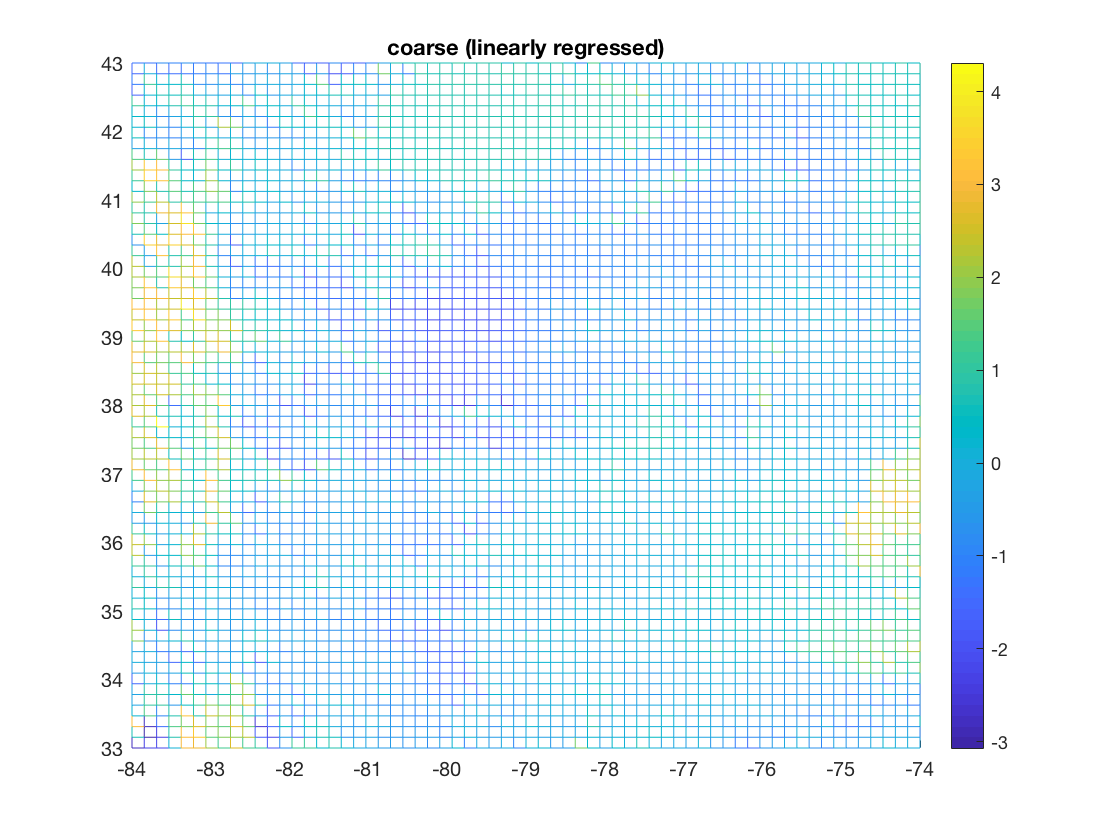}
\includegraphics[width=0.32\textwidth]{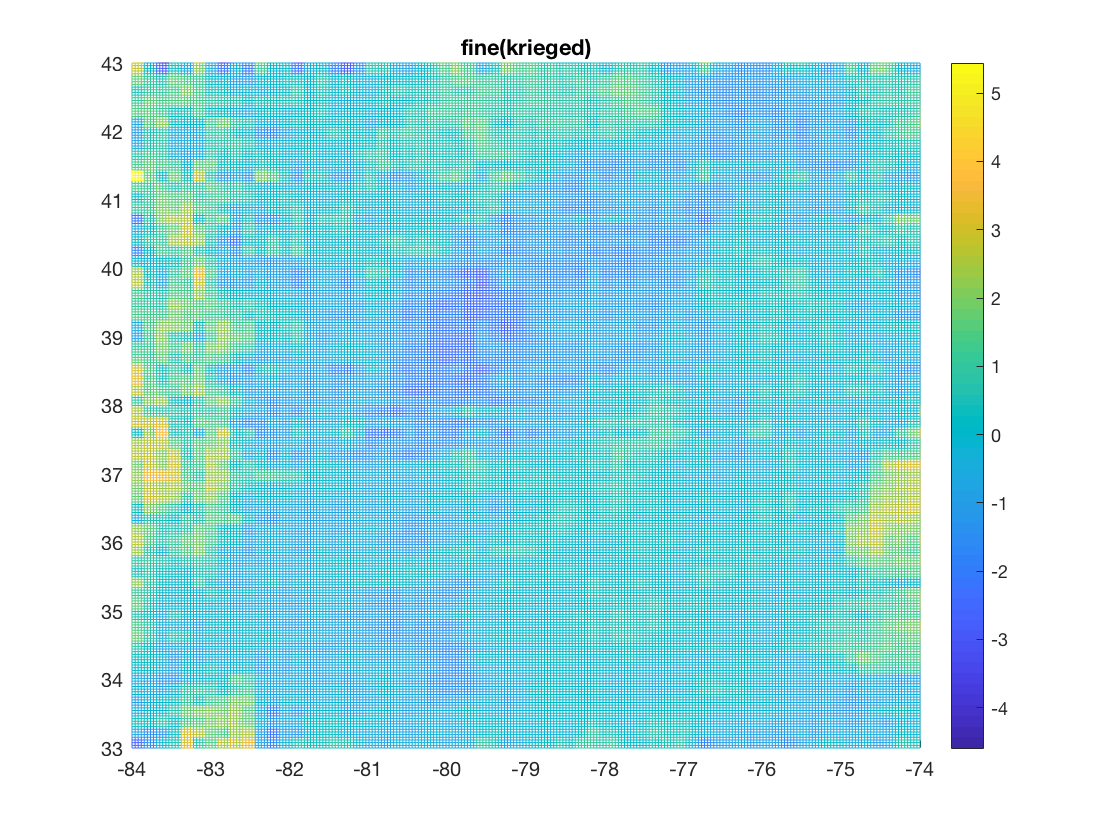}
\caption{(left) 64000 measurements of the moisture; (center) regression on a coarse $65 \times 65$ Cartesian mesh; (right) TT-Kriging approximation on a fine mesh.}
\label{fig:regression_TT}
\end{figure}

The tensor product Kriging is performed as described in Sec. \ref{sec:largeN}.
First, we interpolate the given measurements (Fig.~\ref{fig:regression_TT},~left) onto a (coarse) Cartesian grid with the mesh interval being approximately equal to the average distance between the measurements.
Specifically, we ended up with a $65 \times 65$ grid (Fig.~\ref{fig:regression_TT},~center).
Then the tensor of values on this coarse grid is approximated into a TT decomposition.
Finally, the Kriging estimate \eqref{eq:solve_weights}--\eqref{eq:estimate}
on a fine grid with $257 \times 257$ points (Fig.~\ref{fig:regression_TT},~right) is computed in the TT format using FFT and TT-Cross algorithms.

\subsection{High-dimensional field generation: computational benchmark\label{sub:Performance_Results_time}}
To generate the following 2D, 3D and 10D random fields we used the Matlab script test$\_$generate$\_$y$\_$tt.m in \url{https://github.com/dolgov/TT-FFT-COV}.

\begin{figure}[ht!]
\center
\includegraphics[width=0.9\textwidth]{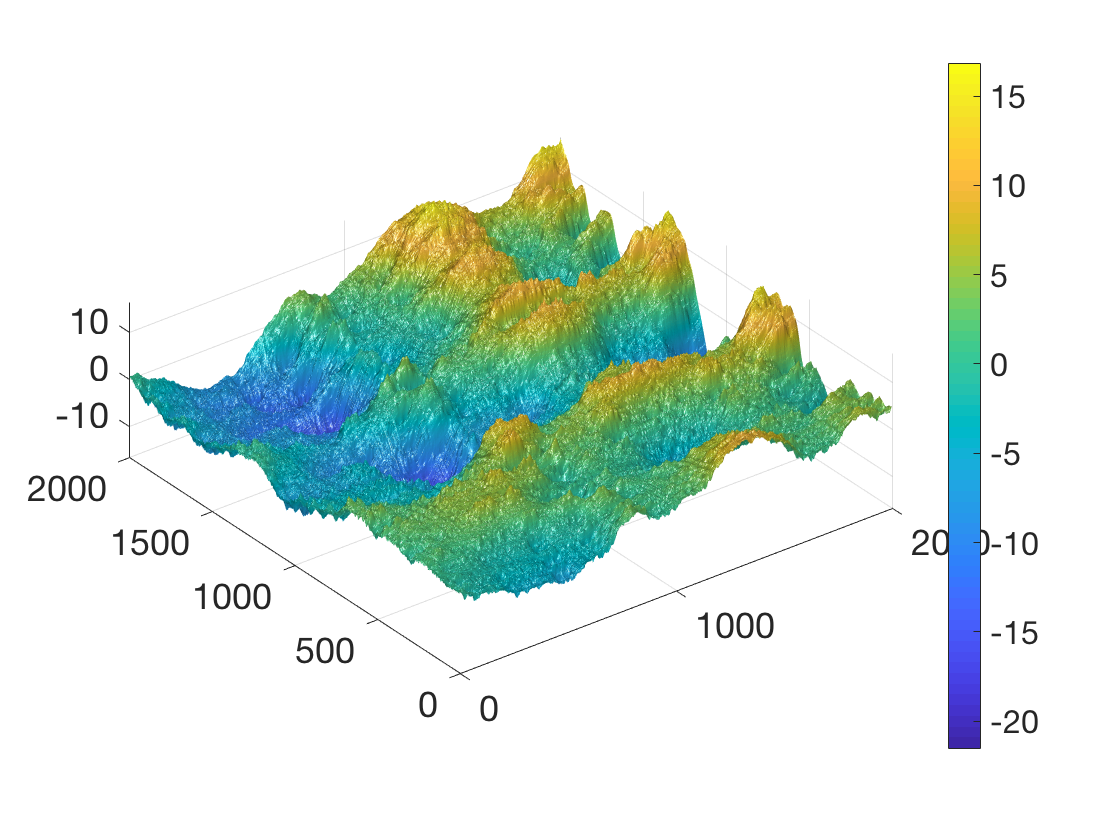}
\caption{High-resolution realization of 2D Mat\'ern random field, computed with TT tensor format in $[0,2000]^2$.}
\label{fig:rv_2D}
\end{figure}

\textbf{2D example.} In this example we generated a high-resolution 2-dimensional Mat\'ern random field in $[0,2000]^2$. One realization is presented in Fig.~\ref{fig:rv_2D}. The smoothness of the Mat\'ern field is $\nu=0.4$, covariance lengths in $x$ and $y$ directions $(1,1)$ and the variance 10. This realization is computed by the following formula in the TT format
\begin{equation}
    \vek {u}'= \mathbf{C}^{1/2}\xib=\sqrt{\frac{1}{n}}\mathbf{F}^\top \mathbf{\Lambda}^{1/2}\xib=\sqrt{\frac{1}{n}}\mathcal{F}^{-1}( {\lambdab}^{1/2}\circ \xib),
\end{equation}
where the inverse Fourier $\mathcal{F}^{-1}$, the square root of eigenvalues $\lambdab^{1/2}$, and tensor product $\xib$ of two Gaussian random vectors are approximated in the TT format. Particularly,  $\xib=\xib_1 \otimes  \xib_2$ is a tensor product of two Gaussian vectors. The size of the first column $\qsse$ of  $\Qsse$ is $3200\times 3600$ and the computing time was 1 sec. With TT procedures one can create very fine resolved random fields in large domains. 
For instance, generation of a random field in the domain $[0,1{.}000{.}000]^2$ with $1{.}600{.}000 \times 1{.}800{.}000$ locations takes less than 1 minute.

\textbf{3D example.} This example is very similar to the previous 2D example. The difference is only that the domain is $[0,100{.000}]^3$ and the size of the first column of $\mathbf{C}$ is $160{.}000\times 180{.}000 \times 160{.}000=4.608\cdot 10^{15} $. The computing time was 3 minutes.

\textbf{10D example.} In this example, we generated a 10-dimensional Mat\'ern random field. One of the dimensions could be time, for example. Table~\ref{table:10dim_details} contains all model parameters and the number of unknowns in (hypothetical) full tensor and in the TT decomposition of the final field $\hat{\mathbf{s}}$.
In this example we computed TT approximation of the first column of the multilevel circulant covariance matrix (cf. \cite{khkaz-conv-2013,KhorToeplCirc17}). Then we diagonalized this circulant matrix via FFT and computed square root of diagonal elements. After that we generated a random field by multiplying the square root with a random vector of the following structure $\xib:=\bigotimes_{\nu=1}^{10} \xib_{\nu}$, where $\xib_{\nu}$ is a normal vector. We note that we never store the whole vector $\xib$ explicitly, but only it's tensor components $\xib_{\nu}$. Also, note that $\xib$ is not Gaussian.

\begin{table}[htbp!]
\centering
\caption{Parameters of the 10-dimensional problem.}
\label{table:10dim_details}  
\begin{footnotesize}
\begin{tabular}{|l|l|}
\hline
 parameter & value  \\ 
\hline
variance of model & 10 \\ \hline
vector of correlation length in $x_1,\ldots,x_{10}$-direction & $[1, 5, 10, 15, 20, 25, 30, 35, 40, 45]$ \\ \hline
length of domain in $x_1,\ldots,x_{10}$-direction & $[10, 50, 100, 150, 200, 250, 300, 350, 400, 450]$\\ \hline
number of elements in $x_1,\ldots,x_{10}$-direction & $[100, 100, 100, 100, 100, 100, 100, 100, 100, 100]$\\ \hline
number of elements in original tensor & $100^{10}=10^{20}$\\ \hline
number of elements in TT tensor & $10^7$ \\ \hline
\end{tabular}
\end{footnotesize}
\end{table}

The TT approximation tolerance is set to $10^{-4}$. 
In the 10-dimensional case above the maximal rank was 143, and the total computing time 118 sec. In the similar 8-dimensional case the maximal rank was 138, and the total computing time 96 sec. Of course, one should observe tensor ranks not only of $\hat{\mathbf{s}}$, but of other steps such as the TT approximation of the measurement vector and of the first column of the covariance matrix. These TT ranks were smaller than the TT ranks of the final solution though.

\section{Discussion and Conclusions\label{sec:Summary-and-Conclusions}}

In this paper, we proposed an FFT-based Kriging that utilizes a low-rank Tensor Train (TT) approximation of the covariance matrix. We apply the TT-Cross algorithm to generate a low-rank decomposition avoiding full tensors which could be well beyond the memory capacity of a desktop PC.

The low-rank format reduces the storage of the embedded circulant covariance matrix from   exponential to linear in the number of variables. The circulant matrix can be diagonalized by FFT. Furthermore, due to the linearity of the Fourier transform, the TT format allows to implement the $d$-dimensional FFT at the cost of $\mathcal{O}(d r^2)$ one-dimensional FFT operations. 

We then use the same technique to generate large Mat\'ern random fields since the diagonalized covariance matrix gives eigen pairs for the spectral expansion of the underlying random field. We show in numerical examples that this method can generate very large random fields with a commonly affordable computational resource.

We demonstrated how to utilize the TT tensor format to speed up such geostatistical tasks as the generation of large random fields, computing kriging coefficients, kriging estimates, conditional covariance, and geostatistical optimal design. 
We used the fact that after discretization on a tensor grid the obtained matrix could be extended to a circulant one. Then, much expensive linear algebra operation could be done via $d$-dimensional FFT. From the definition, one can see that FFT has tensor rank 1. After approximating the first column of the circulant matrix in the TT format (we assumed that such approximation exists) we were able to apply efficient TT tensor arithmetics and speedup expensive calculations even more. Utilizing TT format in FFT calculus allowed us to decrease computational cost and storage from $\mathcal{O}(\Nee\log \Nee)$ to $\mathcal{O}(dr^3\nee)$, where
$r\geq 1$ is the tensor rank,
$d$ the dimensionality of the problem and
$\nee$ is the number of points along the single longest edge of the estimation grid.

The presented numerical techniques have memory requirements
as low as $\mathcal{O}\left(d \nee r^2\right )$. Thus, we achieved log-complexity in the total number of lattice points.
The resulting methods allow much better spatial resolution and significantly reduce the computing time.

The fundamental assumptions are: the covariance matrix is separable or has a TT-rank $r \ll n$, the interpolation grid is a rectangular tensor grid, and the measurements also lie in the tensor grid. The random vector used to generate the random field is a Kronecker product of smaller random vectors.

\section*{Acknowledgments}
The research reported in this publication was supported by funding from the Alexander von Humboldt Foundation. We also would like to thank Wolfgang Nowak for sharing his Matlab code.



\end{document}